\documentclass[aps,showpacs,twocolumn]{revtex4}
\usepackage{epsfig}
\usepackage{amsmath}

\begin{document}
\title{Hidden charmed states and multibody color flux-tube dynamics}

\author{Chengrong Deng$^{a}{\footnote{crdeng@swu.edu.cn}}$,
        Jialun Ping$^b{\footnote{jlping@njnu.edu.cn, corresponding author}}$,
        Hongxia Huang$^b{\footnote{hxhuang@njnu.edu.cn}}$,
       and Fan Wang$^c{\footnote{fgwang@chenwang.nju.edu.cn}}$}
\affiliation{$^a$School of Physical science and technology, Southwest University, Chongqing 400715, China}
\affiliation{$^b$School of Physical science and technology, Nanjing Normal University, Nanjing 210097, China}
\affiliation{$^c$Department of Physics, Nanjing University, Nanjing 210093, China}

\begin{abstract}
Within the framework of the color flux-tube model with a multibody confinement potential, we systematically investigate the hidden charmed states observed in recent years. It can be found that most of them can be described as the compact tetraquark states $[cq][\bar{c}\bar{q}]$ ($q=u,d$ and $s$) in the color flux-tube model. The multibody confinement potential based on the color flux-tube picture is a dynamical mechanism in the formation and decay of the compact tetraquark states.

\end{abstract}


\maketitle

\section{Introduction}

A large amount of hidden charmed $XYZ$ states have been observed by several major particle physics experimental collaborations in the past fifteen years~\cite{pdg}. The discovery of these new hadron states has enriched the charmonium spectroscopy greatly. It is impossible to accommodate all
these states in the conventional $c\bar{c}$-meson framework because the charged $Z_c^+$ states must have a smallest quark component
$c\bar{c}u\bar{d}$ due to carrying one unit charge. The situation provides us an excellent opportunity to deepen our understanding of the complicated non-perturbative behavior of quantum chromodynamics (QCD) in the low energy regime which maybe absent in the traditional $qqq$-baryon
and $q\bar{q}$-meson.

The theoretical physicists have paid much attention to investigate the internal structure of the new hidden charmonium states. Apart from the conventional charmonium states, many exotic candidates, such as meson-meson molecule states, tetra-quark states, charmonium-hybrid states, baryonium states and so on, are proposed in the different theoretical frameworks~\cite{report-chen, rmps}. Even so, the properties of some states
are still not clear so far and therefore more theoretical studies are needed to explain the existing data, which may help us to recognize the mechanism of the low energy non-perturbative behavior better as well as to test various phenomenological models of hadron structure physics.

A systematical investigation on the structures of the hidden charmed states not only is propitious to comprehensively understand the underlying regularities of the properties of the states but also provides new insights into the strong interaction dynamical mechanisms in exotic hadron systems. The goal of the present work is therefore to systematically research the hidden charmed states under the hypothesis of the tetraquark state $[cq][\bar{c}\bar{q}]$ in the color flux-tube model (CFTM) which bases on the lattice QCD (LQCD) picture and the traditional quark models. The model involves a multibody confinement potential instead of a two-body one proportional to the color charge in the traditional quark models~\cite{flux-tube,charged}.

This paper is organized as follows: the CFTM is given in Sec. II. The numerical results and discussions of the hidden charmed states are presented in Sec. III. A brief summary is listed in the last section.

\section{The Color Flux-tube Model}

QCD has been widely accepted as the fundamental theory to describe the interactions among quarks and gluons and the structure of hadrons. However,
it is still difficult for us to derive the hadron spectrum from the QCD directly at present. LQCD was invented to solve QCD numerically through simulations on the lattice, which has been proven very powerful in the calculation of the hadron spectrum and hadron-hadron interactions but so
time-consuming. The phenomenological constituent quark models (CQM) involving the QCD spirits have therefore been applied extensively to study hadron physics. CQM can offer the most complete description of hadron spectra and is probably the most successful phenomenological model of hadron structures~\cite{godfrey}.

The color confinement is the most prominent feature of QCD and should play an essential role in the low energy hadron physics, whose understanding continues to be a challenge in the theoretical physics. A two-body confinement potential proportional to the color charge $\mathbf{\lambda}_i\cdot\mathbf{\lambda}_j$ was introduced to describe quark color confinement in the traditional quark models~\cite{tcqm}, which can well describe the properties of the conventional mesons and baryons. Can the models be directly extended to study multiquark states and hadron-hadron interactions? In fact, the models are well known to be flawed phenomenologically because of the power law van der Waals forces between two color-singlet hadrons~\cite{vandewaals}.

LQCD calculations on mesons, baryons, tetraquark, and pentaquark states reveal color flux-tube or string-like structures~\cite{lqcd1,lqcd2}. The confinement potential of multiquark states is a multibody interaction and can be simulated by a potential proportional to the minimum of the total length of color flux-tubes connecting the quarks and antiquarks to form a multiquark system~\cite{lqcd1,lqcd2}. Based on the traditional quark models and the LQCD picture, the CFTM has been developed in our group~\cite{flux-tube}, in which the confinement potential is a multibody interaction instead of the sum of two-body one in the traditional quark models. In order to simplify the numerical calculation, the linear multibody confinement potential in the LQCD is replaced by a quadratic one. Analytical formulas illustrating the differences between two confinement potentials can be found in
Ref.~\cite{afm}. The numerical comparison studies between linear and quadratic potentials showed that the inaccuracy of this replacement is quite small for the ground states~\cite{qcdben}. In the case of the excited states, the $b\bar{b}$ energy of the $2S$ states in quadratic potential is $\le 1\%$ higher than that in linear potential, around 10\% for $3S$ states, and rise to 20\% for $4S$ states \cite{SNChen}.

The CFTM includes one-gluon-exchange, one-boson-exchange, $\sigma$-meson-exchange and quark confinement potential~\cite{flux-tube}. The model Hamiltonian for the tetraquark states $[cq][\bar{c}\bar{q}]$ is given as follows,
\begin{eqnarray}
H_4 & = & \sum_{i=1}^4 \left(m_i+\frac{\mathbf{p}_i^2}{2m_i}
\right)-T_{C}+\sum_{i>j}^4 V_{ij}+V^C_{min}+V^{C,LS}_{min}, \nonumber\\
V_{ij} & = & V_{ij}^B+V_{ij}^{B,SL}+V_{ij}^{\sigma}+V_{ij}^{\sigma,LS}+V_{ij}^G+V_{ij}^{G,LS}.
\end{eqnarray}
The codes of the quarks (antiquarks) $c$, $q$, $\bar{c}$ and $\bar{q}$ are assumed to be 1, 2, 3 and 4, respectively. Their positions
are denoted as $\mathbf{r}_1$, $\mathbf{r}_2$, $\mathbf{r}_3$ and $\mathbf{r}_4$. $T_{c}$ is the center-of-mass kinetic energy
of the state; $\mathbf{p}_i$ and $m_i$ are the momentum and mass of the $i$-th quark (antiquark), respectively.

The quadratic confinement potential, which is believed to be flavor independent, of the tetraquark state with a diquark-antidiquark
structure has the following form,
\begin{eqnarray}
V^{C}&=&K\left[ (\mathbf{r}_1-\mathbf{y}_{12})^2
+(\mathbf{r}_2-\mathbf{y}_{12})^2+(\mathbf{r}_{3}-\mathbf{y}_{34})^2\right. \nonumber \\
&+&
\left.(\mathbf{r}_4-\mathbf{y}_{34})^2+\kappa_d(\mathbf{y}_{12}-\mathbf{y}_{34})^2\right],
\end{eqnarray}
The variational parameters $\mathbf{y}_{12}$ and $\mathbf{y}_{34}$ are the junctions of two Y-shaped color flux-tube structures. The parameter
$K$ is the stiffness of a three-dimension color flux-tube. The relative stiffness parameter $\kappa_{d}$ for the $d$-dimension compound color flux-tube is $\kappa_{d}=\frac{C_{d}}{C_3}$~\cite{kappa}, where $C_{d}$ is the eigenvalue of the Casimir operator associated with the $SU(3)$
color representation $d$ at either end of the color flux-tube, such as $C_3=\frac{4}{3}$, $C_6=\frac{10}{3}$, and $C_8=3$.

The minimum of the confinement potential $V^C_{min}$ can be obtained by taking the variation of $V^C$ with respect to $\mathbf{y}_{12}$ and
$\mathbf{y}_{34}$, and it can be expressed as
\begin{eqnarray}
V^C_{min}&=& K\left(\mathbf{R}_1^2+\mathbf{R}_2^2+
\frac{\kappa_{d}}{1+\kappa_{d}}\mathbf{R}_3^2\right),
\end{eqnarray}
The canonical coordinates $\mathbf{R}_i$ have the following forms,
\begin{eqnarray}
\mathbf{R}_{1} & = &
\frac{1}{\sqrt{2}}(\mathbf{r}_1-\mathbf{r}_2),~
\mathbf{R}_{2} =  \frac{1}{\sqrt{2}}(\mathbf{r}_3-\mathbf{r}_4), \nonumber \\
\mathbf{R}_{3} & = &\frac{1}{ \sqrt{4}}(\mathbf{r}_1+\mathbf{r}_2
-\mathbf{r}_3-\mathbf{r}_4), \\
\mathbf{R}_{4} & = &\frac{1}{ \sqrt{4}}(\mathbf{r}_1+\mathbf{r}_2
+\mathbf{r}_3+\mathbf{r}_4). \nonumber
\end{eqnarray}
The use of $V^C_{min}$ can be understood here as that the gluon field readjusts immediately to its minimal configuration.

The central parts of one-boson-exchange $V_{ij}^{B}$ and $\sigma$-meson exchange $V_{ij}^{\sigma}$ only take place between light quarks $q$ and $\bar{q}$, while that of one-gluon-exchange $V_{ij}^G$ is universal. $V_{ij}^{B}$, $V_{ij}^{\sigma}$ and $V_{ij}^{G}$ take their standard forms and are listed in the following~\cite{vijande},
\begin{eqnarray}
V_{ij}^{B} & = & V^{\pi}_{ij} \sum_{k=1}^3 \mathbf{F}_i^k
\mathbf{F}_j^k+V^{K}_{ij} \sum_{k=4}^7\mathbf{F}_i^k\mathbf{F}_j^k \nonumber\\
&+&V^{\eta}_{ij} (\mathbf{F}^8_i \mathbf{F}^8_j\cos \theta_P
-\sin \theta_P),\\
V^{\chi}_{ij} & = &
\frac{g^2_{ch}}{4\pi}\frac{m^3_{\chi}}{12m_im_j}
\frac{\Lambda^{2}_{\chi}}{\Lambda^{2}_{\chi}-m_{\chi}^2}
\mathbf{\sigma}_{i}\cdot
\mathbf{\sigma}_{j} \nonumber \\
& \times &\left( Y(m_\chi r_{ij})-
\frac{\Lambda^{3}_{\chi}}{m_{\chi}^3}Y(\Lambda_{\chi} r_{ij})
\right),  \\
V_{ij}^{G} & = & {\frac{\alpha_{s}}{4}}\mathbf{\lambda}^c_{i}
\cdot\mathbf{\lambda}_{j}^c\left({\frac{1}{r_{ij}}}-
{\frac{2\pi\delta(\mathbf{r}_{ij})\mathbf{\sigma}_{i}\cdot
\mathbf{\sigma}_{j}}{3m_im_j}}\right), \\
V^{\sigma}_{ij} & = &-\frac{g^2_{ch}}{4\pi}
\frac{\Lambda^{2}_{\sigma}m_{\sigma}}{\Lambda^{2}_{\sigma}-m_{\sigma}^2}
\left( Y(m_\sigma r_{ij})-
\frac{\Lambda_{\sigma}}{m_{\sigma}}Y(\Lambda_{\sigma}r_{ij})
\right). \nonumber  \\
\end{eqnarray}
where $\chi$ stands for the mesons $\pi$, $K$ and $\eta$, $Y(x)=e^{-x}/x$. $\alpha_s$ is the running strong coupling constant and
takes the following form~\cite{vijande},
\begin{equation}
\alpha_s(\mu_{ij})=\frac{\alpha_0}{\ln\left((\mu_{ij}^{2}+\mu_0^2)/\Lambda_0^2\right)},
\end{equation}
where $\mu_{ij}$ is the reduced mass of two interacting particles.  The function $\delta(\mathbf{r}_{ij})$ in $V_{ij}^G$ should be regularized~\cite{weistein},
\begin{equation}
\delta(\mathbf{r}_{ij})=\frac{1}{4\pi r_{ij}r_0^2(\mu_{ij})}e^{-r_{ij}/r_0(\mu_{ij})},
\end{equation}
where $r_0(\mu_{ij})=\hat{r}_0/\mu_{ij}$. $\Lambda_0$, $\alpha_0$, $\mu_0$ and $\hat{r}_0$ are adjustable model parameters.

The diquark $[cq]$ and antidiquark $[\bar{c}\bar{q}]$ can be considered as compound bosons $\bar{Q}$ and $Q$ with no internal orbital
excitation, and the angular excitations $\mathbf{L}$ are assumed to occur only between $Q$ and $\bar{Q}$ in the present work. In order to facilitate
numerical calculations, the spin-orbit interactions are assumed to take place approximately between compound bosons $\bar{Q}$ and $Q$,
which is consistent with the work~\cite{spin-orbit}. The spin-orbit-related interactions can be expressed as follows
\begin{eqnarray}
V_{\bar{Q}Q}^{G,LS} & \approx& {\frac{\alpha_{s}}{4}}\mathbf{\lambda}^{\bar{c}}_{\bar{Q}}
\cdot\mathbf{\lambda}^c_{Q}{\frac{1}{8M_{\bar{Q}}M_{Q}}}\frac{3}{X^3}
\mathbf{L}\cdot\mathbf{S},\\
V^{\sigma,LS}_{\bar{Q}Q}&\approx&-\frac{g_{ch}^2}{4\pi}
\frac{\Lambda^{2}_{\sigma}}{\Lambda^{2}_{\sigma}-m_{\sigma}^2}
\frac{m_{\sigma}^3}{2M_{\bar{Q}}M_{Q}}\mathbf{L}\cdot\mathbf{S}\nonumber\\
& \times &\left( G(m_\sigma X)-
\frac{\Lambda^3_{\sigma}}{m^3_{\sigma}}G(\Lambda_{\sigma}X)
\right),\\
V_{\bar{Q}Q}^{C,LS} &\approx& \frac{K}{8M_{\bar{Q}}M_{Q}}\frac{\kappa_d}{1+\kappa_d}
\mathbf{L}\cdot\mathbf{S}.
\end{eqnarray}
where the masses of the compound bosons $M_{Q}=M_{\bar{Q}}\approx m_c+m_q$, $X$ is the distance between
the two compound bosons, $G(x)=Y(x)(\frac{1}{x}+\frac{1}{x^2})$, and $S$ stands for the total spin angular
momentum of the tetraquark state $[cq][\bar{c}\bar{q}]$.

It is worth pointing out that the exclusive and most salient feature of the CFTM is that a multibody confinement potential instead of a color dependent two-body one used in the traditional CQM based on the color flux-tube picture in the LQCD is performed to describe multiquark states comparing with other CQM~\cite{vijande,diquonia}. However, the CFTM reduces to the traditional quark model with quadratic confinement potential when it is applied to investigate ordinary hadrons.

The model parameters are determined as follows. The mass parameters $m_{\pi}$, $m_K$ and $m_{\eta}$ in the interaction $V^B_{ij}$ take their experimental values. The cutoff parameters $\Lambda_{\pi}$, $\Lambda_{K}$, $\Lambda_{\eta}$ and $\Lambda_{\sigma}$ and the mixing angle $\theta_{P}$ take the values in the work~\cite{vijande}. The mass parameter $m_{\sigma}$ in the interaction $V_{ij}^{\sigma}$ can be determined through the PCAC relation $m^2_{\sigma}\approx m^2_{\pi}+4m^2_{u,d}$~\cite{masssigma}. The chiral coupling constant $g_{ch}$ can be obtained from the $\pi NN$ coupling constant through
\begin{equation}
\frac{g_{ch}^2}{4\pi}=\left(\frac{3}{5}\right)^2\frac{g_{\pi NN}^2}{4\pi}
\frac{m_{u,d}^2}{m_N^2}.
\end{equation}
The values of the above fixed model parameters are given in Table I. The adjustable parameters and their errors in Table II can be determined by fitting the masses of the ground states of mesons in Table III using Minuit program. Once the meson masses are obtained, one can calculate the threshold energies of the tetraquark states $[cq][\bar{c}\bar{q}]$ simply by adding the masses of two mesons in Table III.

\begin{table}[ht]
\caption{Fixed model parameters.}
\begin{tabular}{cccccccccccccc}
\hline
Para.        &  Valu. & Unit      &~~~&  Para.              & Valu.& Unit      &~~~& Para.                   & Vale.             & Unit      \\
$m_{ud}$     &  280   & MeV       &   &  $m_{\sigma}$       & 2.92 & fm$^{-1}$ &   & $\Lambda_{\eta}$        &  5.2              & fm$^{-1}$ \\
$m_{\pi}$    &  0.7   & fm$^{-1}$ &   &  $\Lambda_{\pi}$    & 4.2  & fm$^{-1}$ &   & $\theta_P$              & $-\frac{\pi}{12}$ & ...       \\
$m_{K}$      &  2.51  & fm$^{-1}$ &   &  $\Lambda_{\sigma}$ & 4.2  & fm$^{-1}$ &   & $\frac{g^2_{ch}}{4\pi}$ &  0.43             & ...       \\
$m_{\eta}$   &  2.77  & fm$^{-1}$ &   &  $\Lambda_{K}$      & 5.2  & fm$^{-1}$           \\
\hline
\end{tabular}
\caption{Adjustable model parameters.}
\begin{tabular}{ccccccccccc}
\hline
Para.        &   $x_i\pm\Delta x_i$  & Unit   &     Para.       &   $x_i\pm\Delta x_i$  &  Unit         \\
$m_{s}$      &   $511.78\pm0.228$    & MeV    &    $\alpha_0$   &   $4.554\pm0.018$     &   ...         \\
$m_{c}$      &   $1601.7\pm0.441$    & MeV    &    $K$          &   $217.50\pm0.230$    &  MeV$\cdot$fm$^{-2}$         \\
$m_{b}$      &   $4936.2\pm0.451$    & MeV    &    $\mu_0$      &   $0.0004\pm0.540$    &  MeV          \\
$\Lambda_0$  &   $9.173\pm0.175$     & MeV    &    $r_0$        &   $35.06\pm0.156$     &  MeV$\cdot$fm \\
\hline
\end{tabular}
\caption{Ground state meson spectra, unit in MeV.\label{meson}}
\begin{tabular}{ccccccccccccc}
\hline
States         &  ~~~$E_2\pm\Delta E_2$ &   ~~PDG~~    &     States         &  ~~~$E_2\pm\Delta E_2$  &    ~~PDG~~  \\
$\pi$          &     $142\pm26$         &     139      &     $\eta_c$       &     $2912\pm5$          &     2980    \\
$K$            &     $492\pm20$         &     496      &     $J/\Psi$       &     $3102\pm4$          &     3097    \\
$\rho$         &     $826\pm4$          &     775      &     $B^0$          &     $5259\pm5$          &     5280    \\
$\omega$       &     $780\pm4$          &     783      &     $B^*$          &     $5301\pm4$          &     5325    \\
$K^*$          &     $974\pm4$          &     892      &     $B_s^0$        &     $5377\pm5$          &     5366    \\
$\phi$         &     $1112\pm4$         &     1020     &     $B_s^*$        &     $5430\pm4$          &     5416    \\
$D^{\pm}$      &     $1867\pm8$         &     1880     &     $B_c$          &     $6261\pm7$          &     6277    \\
$D^*$          &     $2002\pm4$         &     2007     &     $B_c^*$        &     $6357\pm4$          &     ...     \\
$D_s^{\pm}$    &     $1972\pm9$         &     1968     &     $\eta_b$       &     $9441\pm8$          &     9391    \\
$D_s^*$        &     $2140\pm4$         &     2112     &     $\Upsilon(1S)$ &     $9546\pm5$          &     9460    \\

\hline

\end{tabular}
\end{table}

\section{numerical results and discussions}

Within the framework of the diquark-antidiquark configuration, the wave function of the state $[cq][\bar{c}\bar{q}]$ can be written as a sum
of the following direct products of color $\chi_c$, isospin $\eta_i$, spin $\chi_s$ and spatial
$\phi$ terms,
\begin{eqnarray}
\Phi^{[cq][\bar{c}\bar{q}]}_{IM_IJM_J} &=&
\sum_{\alpha}\xi_{\alpha}\left[ \left[
\left[\phi_{l_am_a}^G(\mathbf{r})\chi_{s_a}\right]^{[cq]}_{s_a}
\left[\phi_{l_bm_b}^G(\mathbf{R})\right.\right.\right.\nonumber\\
& \times & \left.\left.\left.\chi_{s_b}\right]^{[\bar{c}\bar{q}]}_{s_b}
\right ]_{S}^{[cq][\bar{c}\bar{q}]}
F_{LM}(\mathbf{X})\right]^{[cq][\bar{c}\bar{q}]}_{JM_J}\\
& \times &
\left[\eta_{i_a}^{[cq]}\eta_{i_b}^{[\bar{c}\bar{q}]}\right]_{IM_I}^{[cq][\bar{c}\bar{q}]}
\left[\chi_{c_a}^{[cq]}\chi_{c_b}^{[\bar{c}\bar{q}]}\right]_{CW_C}^{[cq][\bar{c}\bar{q}]},
\nonumber
\end{eqnarray}
In which the subscripts $a$ and $b$ represent the diquark $[cq]$ and antidiquark $[\bar{c}\bar{q}]$, respectively. The parity of the states is
related to the angular excitations $\mathbf{L}$ between $Q$ and $\bar{Q}$ as $P=(-1)^L$ because of no internal orbital excitation in the $Q$ and $\bar{Q}$. Considering a pair of charge-conjugated bosons $Q\bar{Q}$, we can obtain the $C$-parity $C=(-1)^{L+S-s_a-s_b}$ because the total wavefunction has to be completely symmetric under exchange of coordinates and spin of the bosons $Q$ and $\bar{Q}$.

The relative spatial coordinates $\mathbf{r}$, $\mathbf{R}$ and $\mathbf{X}$ can be defined as,
\begin{eqnarray}
\mathbf{r}&=&\mathbf{r}_1-\mathbf{r}_2,~~~\mathbf{R}=\mathbf{r}_3-\mathbf{r}_4 \nonumber\\
\mathbf{X}&=&\frac{m_1\mathbf{r}_1+m_2\mathbf{r}_2}{m_1+m_2}-\frac{m_3\mathbf{r}_3+m_4\mathbf{r}_4}{m_3+m_4}.
\end{eqnarray}
It is worth mentioning that this set of coordinate is just one possible choice of many coordinates and however most propitious to describe the correlation of two quarks in the diquark. In order to obtain a reliable solution, a high precision numerical method is indispensable. The Gaussian Expansion Method(GEM)~\cite{GEM}, which has been proven to be rather powerful to solve few-body problem, is therefore  used to study four-quark systems in present work. According to the GEM, three relative motion wave functions can be written as,
\begin{eqnarray}
\phi^G_{l_am_a}(\mathbf{r})=\sum_{n_a=1}^{n_{amax}}c_{n_a}N_{n_al_a}r^{l_a}e^{-\nu_{n_a}r^2}Y_{l_am_a}(\hat{\mathbf{r}})\nonumber\\
\psi^G_{l_bm_b}(\mathbf{R})=\sum_{n_b=1}^{n_{bmax}}c_{n_b}N_{n_bl_b}R^{l_b}e^{-\nu_{n_b}R^2}Y_{l_bm_b}(\hat{\mathbf{R}})\nonumber\\
\chi^G_{LM}(\mathbf{X})=\sum_{n_c=1}^{n_{cmax}}c_{n_c}N_{LM}X^{L}e^{-\nu_{n_c}X^2}Y_{LM}(\hat{\mathbf{X}})\nonumber
\end{eqnarray}
More details of the relative motion wave functions can be found in the paper~\cite{GEM}.

The color representation of the diquark maybe symmetrical $[cq]_{\bar{\mathbf{3}}_c}$ or antisymmetrical $[cq]_{\mathbf{6}_c}$, whereas that of the antidiquark maybe symmetrical $[\bar{c}\bar{q}]_{\mathbf{3}_c}$ or antisymmetrical ${\mathbf{6}_c}$. 
 As the tetraquark states must be a overall color singlet, the possible diquark-antidiquark color combinations only have two ways according to color coupling rule: $\left [ [cq]_{\bar{\mathbf{3}}_c}\otimes[\bar{c}\bar{q}]_{\mathbf{3}_c}\right ]_{\mathbf{1}}$ and $\left [[cq]_{\mathbf{6}_c}\otimes[\bar{c}\bar{q}]_{\bar{\mathbf{6}}_c}\right ]_{\mathbf{1}}$, which are named ``true" state and ``mock" state~\cite{diquonia}, respectively. A real physical state should be their mixture because of the coupling between two states. The total spin wave function can be written as $S=s_a\oplus s_b$. Then we have the following basis vectors as a function of the total spin $S$,
\begin{eqnarray}
S&=&0: ~~0\oplus0,~~1\oplus1\nonumber\\
S&=&1: ~~0\oplus1,~~1\oplus0,~~1\oplus1\nonumber\\
S&=&2: ~~1\oplus1\nonumber
\end{eqnarray}

With respect to the flavor wavefunction, we only consider $SU_f(2)$ symmetry in the present work. The quarks $s$ and $c$ have isospin zero so that they do not contribute to the total isospin. The flavor wave functions of the states only consisting of $u$ and $d$ quarks and their anti-particles are similar to those of spin.

The converged numerical results can be obtained by solving the four-body Schr\"{o}dinger equation
\begin{eqnarray}
(H_4-E_4)\Phi^{[cq][\bar{c}\bar{q}]}_{IM_IJM_J}=0.
\end{eqnarray}
with the Rayleigh-Ritz variational principle. The energies $E_4\pm\Delta E_4$ of the states $[cq][\bar{c}\bar{q}]$ with $n^{2S+1}L_J$ and
$IJ^{PC}$ or $IJ^{P}$ which maybe consistent with the quantum numbers of the experimental states are systematically calculated and presented
in Table IV.

Next, we discuss the properties of the hidden charmed states observed in experiments and their possible candidates in the CFTM. The state $X(3872)$ was first discovered in the hidden charmed family and favors $IJ^{PC}=01^{++}$~\cite{x3872}. Various interpretations have been proposed to explain its structure in the different theoretical framework since 2003~\cite{report-chen}, such as diquark-antidiquark state and molecule state. However, the property of the state has been not fully understood so far. In the CFTM, the tetraquark $[cq][\bar{c}\bar{q}]$ with $J^{PC}=01^{++}$ and $1^3S_1$ has a mass of $3926\pm9$ MeV, which is a little higher than the experimental data although a four-body confinement potential instead of two-body one was applied~\cite{x3872slamanca}. The state $Z(3930)$ was reported by the Belle Collaboration in 2005 and favors the $IJ^{PC}=02^{++}$ assignment~\cite{z3930}. The state $X(3915)$ was first reported by the Belle Collaboration and then confirmed by the BarBar Collaboration, the spin-parity of the state $J^P=0^+$ was favored~\cite{x3915,x3915confirmed}. The states $Z(3930)$ and $X(3915)$ were suggested as the good candidates of the $P$-wave charmonia, $\chi^{\prime}_{c0}(2P)$ and $\chi^{\prime}_{c2}(2P)$~\cite{pdg,z3930}, respectively. There is a big gap between the energies of the tetraquark states with $J^P=2^+$ and $0^+$ in the CFTM and those of the states $Z(3930)$ and $X(3915)$, respectively.

\begin{table*}
\caption{The hidden charmonium states observed in experiments and the energy $E_4\pm\Delta E_4$ of the tetraquark states with $IJ^{PC}$ or
$IJ^{P}$ and $n^{2S+1}L_J$ in the CFTM, where $q'$ stands for $u$ and $d$ quarks, unit in MeV.}
\begin{tabular}{cccc||ccccc||cccccccccccc}

\hline\hline
       &~~~~Experiments~~~~&&&&& CFTM &&&\\
States &  Mass  & Width  &$IJ^{PC}/J^P$ &~~Flavors~~&  $IJ^{PC}/J^P$ & $n^{2S+1}L_J$ & $E_4\pm\Delta E_4$ &~$T_{M_1M_2}$, $E_b$~& Consistency  \\
$X(3872)$&$3871.69\pm0.17$             &$<1.2$                       &$01^{++}$ &$[cq'][\bar{c}\bar{q}']$&$01^{++}$& $1^3S_1$&$3926\pm9$ &
$D\bar{D}^*$,   57&$\times$\\
$X(3823)$&$3821.3\pm1.3\pm0.3$         &$<16$                        &$02^{--}$ &          ...           &   ...   &     ... &     ...   &
       ...        &$\times$\\
$Z(3930)$&$3929\pm5\pm2$               &$29\pm10\pm2$                &$02^{++}$ &$[cq'][\bar{c}\bar{q}']$&$02^{++}$& $1^5S_0$&$3984\pm7$ &
$D^*\bar{D}^*$, 20&$\times$\\
$X(3915)$&$3915\pm3\pm2$               &$17\pm10\pm3$                &$00^{++}$ &$[cq'][\bar{c}\bar{q}']$&$00^{++}$& $1^1S_0$&$3859\pm10$&
$D\bar{D}$, 125&$\times$   \\
$X(3940)$&$3942^{+7}_{-6}\pm6$         &$31^{+10}_{-8}\pm5$          &$??^{?+}$ &$[cq'][\bar{c}\bar{q}']$&$01^{++}$& $2^3S_0$&$3936\pm9$ &
$D\bar{D}^*$, 47&$\surd$   \\
$X(4160)$&$4156^{+25}_{-20}\pm15$      &$139^{+111}_{-61}\pm21$      &$??^{?+}$ &$[cq'][\bar{c}\bar{q}']$&$01^{--}$& $1^5P_1$&$4140\pm8$ &
$D^*\bar{D}^*$, 136&$\surd$\\
$Y(3940)$&$3919.1^{+3.8}_{-3.5}\pm2.0$ &$37^{+26}_{-15}\pm8$         &$??^{?+}$ &$[cq'][\bar{c}\bar{q}']$&$01^{++}$& $1^3S_1$&$3926\pm9$ &
$D\bar{D}^*$, 57&$\surd$   \\
$X(4350)$&$4350.6^{+4.6}_{-5.1}\pm0.5$ &$13^{+18}_{-9}\pm4$          &$0^+,2^+$ &$[cq'][\bar{c}\bar{q}']$&$02^{++}$& $2^5D_2$&$4375\pm9$ &
$D^*\bar{D}^*$, 371& $\surd$ \\
$X(4140)$&$4146.5\pm4.5^{+4.6}_{-2.8}$ &$83\pm21^{+21}_{-14}$        &$01^{++}$ &         ...            &     ... &     ... &     ...   &
      ...        & $\times$\\
$X(4274)$&$4273.3\pm8.3^{+17.2}_{-3.6}$&$56\pm11^{+8}_{-11}$         &$01^{++}$ &$[cs][\bar{c}\bar{s}]$  &$01^{+}$ & $1^3S_1$&$4259\pm8$ &
$D_s\bar{D}_s^*$, 147& $\surd$ \\
$X(4500)$&$4506\pm11^{+12}_{-15}$      &$92\pm21^{+21}_{-20}$        &$00^{++}$ &$[cs][\bar{c}\bar{s}]$  &$00^{+}$ & $1^5D_0$&$4596\pm7$ &
$D_s^*\bar{D}_s^*$, 316& $\surd$ \\
$X(4700)$&$4704\pm10^{+14}_{-24}$      &$120\pm31^{+42}_{-33}$       &$00^{++}$ &$[cs][\bar{c}\bar{s}]$  &$00^{+}$ & $2^5D_0$&$4704\pm7$ &
$D_s^*\bar{D}_s^*$, 424& $\surd$ \\
$Y(4008)$&$4008\pm40^{+114}_{-28}$     &$226\pm44\pm87$              &$1^-$     &$[cq'][\bar{c}\bar{q}']$&$01^{--}$& $1^1P_1$&$4076\pm8$ &
$D\bar{D}$, 342& $\surd$ \\
$Y(4260)$&$4251\pm9$                   &$120\pm12$                   &$01^{--}$ &$[cq'][\bar{c}\bar{q}']$&$01^{--}$& $2^5P_1$&$4254\pm8$ &
$D^*\bar{D}^*$, 250& $\surd$ \\
$Y(4360)$&$4354\pm10$                  &$78\pm16$                    &$01^{--}$ &$[cq'][\bar{c}\bar{q}']$&$01^{--}$& $1^5F_1$&$4384\pm8$ &
$D^*\bar{D}^*$, 380& $\surd$ \\
$Y(4220)$&$4218.4^{+5.5}_{-4.5}\pm0.9$ &$66.0^{+12.3}_{-8.3}\pm0.4$  &$01^{--}$ &$[cq'][\bar{c}\bar{q}']$&$01^{--}$& $2^5P_1$&$4254\pm8$ &
$D^*\bar{D}^*$, 250& $\surd$ \\
$Y(4390)$&$4391.5^{+6.3}_{-6.8}\pm1.8$ &$139.5^{+16.2}_{-20.6}\pm0.6$&$01^{--}$ &$[cq'][\bar{c}\bar{q}']$&$01^{--}$& $1^5F_1$&$4384\pm8$ &
$D^*\bar{D}^*$, 380& $\surd$ \\
$Y(4660)$  &$4665.3\pm10$              &$53\pm16$                    &$01^{--}$ &$[cq'][\bar{c}\bar{q}']$&$01^{--}$& $3^5F_1$&$4610\pm8$ &
$D^*\bar{D}^*$, 606& $\surd$ \\
$Y(4630)$  &$4634^{+8+5}_{-7-8}$       &$92^{+40+10}_{-24-21}$       &$01^{--}$ &$[cq'][\bar{c}\bar{q}']$&$01^{--}$& $3^5F_1$&$4610\pm8$ &
$D^*\bar{D}^*$, 606& $\surd$ \\
$Z_c^+(3900)$&$3888.7\pm3.4$               &$35\pm7$                 &$11^{+}$  &$[cu][\bar{c}\bar{d}]$  &$11^{+}$ & $1^3S_1$&$3858\pm10$&
$D\bar{D}^*$, $-11$& $\surd$ \\
$Z_c^+(3885)$&$3883.9\pm1.5\pm4.2$         &$24.8\pm3.3\pm11.0$      &$11^{+}$  &$[cu][\bar{c}\bar{d}]$  &$11^{+}$ & $1^3S_1$&$3858\pm10$&
$D\bar{D}^*$, $-11$& $\surd$ \\
$Z_c^+(4020)$&$4022.9\pm0.8\pm2.7$         &$7.9\pm2.7\pm2.6$        &$??^?$    &$[cu][\bar{c}\bar{d}]$  &$12^{+}$ &$1^5S_2$ &$4001\pm7$ &
$D^*\bar{D}^*$, $-3$& $\surd$ \\
$Z_c^+(4025)$&$4026.3\pm2.6\pm3.7$         &$24.8\pm5.6\pm7.7$       &$??^?$    &$[cu][\bar{c}\bar{d}]$  &$12^{+}$ &$1^5S_2$ &$4001\pm7$ &
$D^*\bar{D}^*$, $-3$& $\surd$ \\
$Z_c^+(4051)$&$4051.3\pm14^{+20}_{-41}$    &$82^{+21+47}_{-17-22}$   &$??^?$    &$[cu][\bar{c}\bar{d}]$  &$11^{-}$ &$1^1P_1$ &$4075\pm8$ &
$D\bar{D}^*$, 206& $\surd$ \\
$Z_c^+(4248)$&$4248.3^{+44+180}_{-29-35}$  &$177^{+54+316}_{-39-61}$ &$??^?$    &$[cu][\bar{c}\bar{d}]$  &$11^{+}$ &$1^5D_1$ &$4273\pm7$ &
$D^*\bar{D}^*$, 269& $\surd$ \\
$Z_c^+(4200)$&$4196.9^{+31+17}_{-29-13}$   &$370^{+70+70}_{-70-132}$ &$11^+$    &$[cu][\bar{c}\bar{d}]$  &$11^{+}$ &$1^3D_1$ &$4235\pm7$ &
$D\bar{D}^*$, 366& $\surd$ \\
$Z_c^+(4240)$&$4239.3\pm18^{+45}_{-10}$    &$220\pm47^{+108}_{-74}$  &$??^?$    &$[cu][\bar{c}\bar{d}]$  &$11^{+}$ &$1^5D_1$ &$4273\pm7$ &
$D^*\bar{D}^*$, 269& $\surd$ \\
$Z_c^+(4430)$&$4478\pm40^{+15}_{-18}$      &$181\pm31$               &$1^{+}$   &$[cu][\bar{c}\bar{d}]$  &$11^{+}$ &$3^5D_1$ &$4497\pm7$ &
$D^*\bar{D}^*$, 493& $\surd$ \\
$Z_c^+(4475)$&$4475\pm^{+15}_{-25}$        &$172\pm13^{+37}_{-34}$   &$1^{+}$   &$[cu][\bar{c}\bar{d}]$  &$11^{+}$ &$3^5D_1$ &$4497\pm7$ &
$D^*\bar{D}^*$, 493& $\surd$ \\
\hline\hline

\end{tabular}
\end{table*}

The states $X(3940)$ and $X(4160)$ were discoverd in the double charmonium production $e^+e^-\rightarrow J/\Psi X$~\cite{x3940&4160}. Many work described the state $X(3940)$ as the $\eta_c(3S)$ charmonium state. However, a problem of the description of the $X(3940)$ is that its mass is
a bit lower than theoretical prediction~\cite{highercharm1,highercharm2}. The energy of the state $[cq][\bar{c}\bar{q}]$ with $J^{PC}=0^{++}$ and $2^3S_1$ is $3936\pm9$ in the CFTM, which is highly consistent with the experimental data of the state $X(3940)$. The main component of the state $X(3940)$ maybe therefore the tetraquark state. With respect to the state $X(4160)$, the CFTM can describes it as the tetraquark state $[cq][\bar{c}\bar{q}]$ with $J^{PC}=1^{--}$. These assignments are different from the results of the QCD sum rules approach, where the masses of the tetraquark states $[cq][\bar{c}\bar{q}]$ are much higher than the masses of the X(3940) and X(4160) and does not support them to be charmonium-like tetraquark states~\cite{czcharm}.

The state $X(3823)$ was observed by the Belle Collaboration and suggested $J^{PC}=2^{--}$~\cite{x3823}. The energy of the tetrequark state $[cq][\bar{c}\bar{q}]$ with negative parity $(L=1,3)$ is much higher than that of the state $X(3823)$ (see the Table IV). In this way, the state
can not be described as a tetraquark state in the CFTM.

The state $X(4350)$ was reported by the Belle Collaboration in the process of $\gamma\gamma\rightarrow J/\psi\phi $, its quantum number is either $J^{P}=0^{+}$ or $2^{+}$~\cite{x4350}. The state was describe as the charmonium state $\chi_{c2}(3P)$ in Refs.~\cite{highercharm1,x4350charmonium}. Furthermore, QCD sum rules disfavored the assignment of the X(4350) as the exotic charmonium-like tetraquark or molecular state~\cite{x4350qcdsum}. However, the state $X(4350)$ can be interpreted as the $2^5D_2$ and $2^+$ tetraquark state $[cq][\bar{c}\bar{q}]$ with positive parity in the CFTM.

The states $X(4140)$ and $X(4274)$ were first reported by the CDF Collaboration~\cite{x4140,x4274}. Recently, the LHCb Collaboration confirmed
the two states in the $J/\psi \phi$ invariant mass distribution and determined their spin-parity both to be $J^P=1^+$~\cite{x4500x4700}. At the same time the two states $X(4500)$ and $X(4700)$ with $J^{P}=0^+$ states were observed in the $J/\psi \phi$ invariant mass distribution~\cite{x4500x4700}. These four states in the $J/\psi \phi$ invariant mass spectrum attracted much attention because they may contain both a $c\bar{c}$ pair and an $s\bar{s}$ pair, which implies that they may be exotic states. The tetraquark states $[cs][\bar{c}\bar{s}]$ with $J^{PC}=1^{++}$ and $0^{++}$ are investigated in the CFTM. The lowest energy of the state with $J^{PC}=1^{++}$ is $4259\pm8$ MeV, which is much higher than that of the state $X(4140)$ but highly consistent with the state $X(4270)$. The tetraquark state $[cs][\bar{c}\bar{s}]$ with $J^{PC}=0^{++}$ and $1^5D_0$ has a mass of $4596\pm7$ MeV, which is a little higher than that of the state $X(4500)$. However, we cannot rule out the possibility of interpreting the state $X(4500)$ as the $[cs][\bar{c}\bar{s}]$ with $J^{PC}=0^{++}$ and $1^5D_0$. The energy of the first radial excited state $[cs][\bar{c}\bar{s}]$ with $2^5D_0$ is $4704\pm7$ MeV, which is in full accord with that of the state $X(4700)$. In brief, the states $X(4274)$, $X(4500)$ and $X(4700)$ can be described as the compact tetraquark states $[cs][\bar{c}\bar{s}]$ in the CFTM, which is supported by the work~\cite{cscs}. However, the state $X(4140)$ cannot be accommodated in the CFTM.

The $Y$-states, $Y(4260)$, $Y(4008)$, $Y(4360)$, $Y(4220)$, $Y(4390)$, $Y(4630)$ and $Y(4660)$, are vector meson states with $J^{PC}=1^{--}$
because they were produced from the $e^+e^-$ annihilation. The state $Y(4260)$ was first observed by the BaBar Collaboration in 2005~\cite{y4260}.
Later, it was confirmed by both the CLEO and Belle collaborations in the same process~\cite{y4260confirmed}. Many pictures were proposed to
describe its internal structure~\cite{report-chen}, such as charmonium states, hybrid charmonium, tetraquark state and molecular state. The
energies of the state $[cq][\bar{c}\bar{q}]$ with $J^{PC}=1^{--}$ and $2^5P_1$, $4254\pm8$ MeV, is extremely in agreement with that of the state $Y(4260)$ in the CFTM. The interpretation of the $Y(4260)$ as the tetraquark state $[cq][\bar{c}\bar{q}]$ is consistent with the conclusion in
the work~\cite{y4260diquark}. The state $Y(4008)$ was reported and confirmed by the Belle Collaboration~\cite{y4008}. The energy of the state $[cq][\bar{c}\bar{q}]$ with $J^{PC}=1^{--}$ and $1^{1}P_1$ is $4076\pm8$ MeV, which is a little higher than the central value of the state
$Y(4008)$. However, it is still within the error-bar of the state $Y(4008)$. The state $Y(4360)$ was observed by the BaBar Collaboration
and confirmed by the Belle Collaboration~\cite{y4360, y4360confirmed}. The tetraquark state $[cq][\bar{c}\bar{q}]$ with $J^{PC}=1^{--}$ and $1^5F_1$ has a energy of $4384\pm8$ MeV, which is very close to that of the state $Y(4360)$. The states $Y(4220)$ and $Y(4390)$ were observed in the $e^+e^-\rightarrow \pi^+\pi^-h_c$ cross sections around 4.22 GeV and 4.39 GeV, respectively \cite{y4220&4390}. The energies of the states $Y(4220)$ and $Y(4390)$ are also close to those, $4254\pm8$ MeV and  $4384\pm8$ MeV, of the state $[cq][\bar{c}\bar{q}]$ with $2^5P_1$ and $1^5F_1$ in the CFTM, respectively. The state $Y(4660)$ was observed in the initial-state radiation process $e^+e^-\rightarrow \gamma_{ISR}Y(4660)$~\cite{y4360confirmed}. The state $Y(4630)$ was observed in the exclusive $e^+e^-\rightarrow \Lambda_c\Lambda_c^-$ cross section~\cite{y4630}. The masses and widths of the two states are consistent with each other within errors~\cite{y4630}. The two states therefore may be the same state or structure~\cite{y4630&y4660}.
The energy of the tetraquark state $[cq][\bar{c}\bar{q}]$ with $J^{PC}=1^{--}$ and $3^5F_1$ has a mass of $4610\pm8$ MeV, which is very close to the experimental values of the states $Y(4630)$ and $Y(4660)$. In this way, it is favored that the main components of the states $Y(4008)$, $Y(4260)$, $Y(4360)$, $Y(4220)$, $Y(4390)$, $Y(4630)$ and $Y(4660)$ can be described as the tetraquark states $[cq][\bar{c}\bar{q}]$ in the CFTM.

A great deal of charged charmonium-like $Z^+_c$-states have been reported until now. The BESIII Collaboration reported the four states $Z^+_c(3900)$, $Z^+_c(3885)$, $Z^+_c(4020)$ and $Z^+_c(4025)$~\cite{zc3900,zc3885,zc4020,zc4025}. The state $Z^+_c(3900)$ may correspond to the same state as the state $Z^+_c(3885)$ with $J^P=1^+$ because of the similar mass and width~\cite{zc3885}. The charged state $[cu][\bar{c}\bar{d}]$ with $J^P=1^+$ and $1^3S_1$ has a mass of $3858\pm10$ MeV in the CFTM, which is very close to those of the two charged states $Z^+_c(3885)$ and $Z^+_c(3900)$. It cannot be excluded that the main component of the two states $Z^+_c(3885)$ and $Z^+_c(3900)$ is the state $[cu][\bar{c}\bar{d}]$ with $J^P=1^+$ and $1^3S_1$, which is supported by many theoretical work~\cite{zc3900tetraquark}. The pair $Z^+_c(4020)$ and $Z^+_c(4025)$ might also be the same state due to the similar mass. However, the spin and parity of two states have been unclear so far. QCD sum rule identified the states $Z^+_c(4020)$ and $Z^+_c(4025)$ as a tetraquark state $[cu][\bar{c}\bar{d}]$ with $J^P=1^+$~\cite{zgwang}, the same approach also favored a tetraquark state but with different quantum numbers $J^P=2^+$ and $1^5S_2$~\cite{sumrule2+}. In the CFTM, the nearest tetraquark state $[cu][\bar{c}\bar{d}]$ to the $Z^+_c(4020)$ and $Z^+_c(4025)$ occupies quantum numbers $J^P=2^+$ and $1^5S_2$.

The Belle Collaboration reported the states $Z^+_1(4050)$ and $Z^+_2(4250)$ in the $\pi^+\chi_{c1}$ invariant mass distribution near 4.1 GeV in exclusive $\bar{B}^0\rightarrow K^-\pi^+\chi_{c1}$ decays~\cite{zc4020&4250}. Many theoretical researches do not favor the molecular assignment
of the states $Z^+_1(4050)$ and $Z^+_2(4250)$~\cite{zc4200nom}. The tetraquark states $[cu][\bar{c}\bar{d}]$ with $J^P=1^-$ and $1^1P_1$ and
$J^P=1^+$ and $1^5D_1$ have the energies of $4075\pm8$ MeV and $4273\pm7$ MeV in the CFTM, respectively, which are consistent with those of $Z^+_1(4050)$ and $Z^+_2(4250)$, respectively. The states $Z^+_1(4050)$ and $Z^+_2(4250)$ may therefore be assigned as the tetraquark states $[cu][\bar{c}\bar{d}]$ with $J^P=1^-$ and $1^1P_1$ and $J^P=1^+$ and $1^5D_1$, respectively, in the CFTM. The state $Z^+_c(4200)$ was reported
by the Belle Collaboration in the amplitude analysis of $\bar{B}_0\rightarrow J/\psi K^-\pi^+$ decays, which decays into $J\psi \pi^+$ and prefers $1^+$~\cite{zc4200}. The state $Z_c^+(4200)$ can be described as the tetraquark state $[cu][\bar{c}\bar{d}]$ with $J^P=1^+$ and $1^3D_1$ in the
CFTM. The study of the three-point function sum rules on this state supports the tetraquark interpretation~\cite{wchen4200}.

The state $Z^+_c(4430)$ was observed in the $\pi^{\pm}\psi^{\prime}$ invariant mass distribution near 4.43 GeV in the $B\rightarrow K\pi^{\pm}\psi^{\prime}$ decays~\cite{zc4430}. The $J^P$ of the state $Z^+_c(4430)$ is determined unambiguously to be $1^+$~\cite{zc4430confirmed}.
The state $Z^+_c(4475)$ was discovered by the Belle Collaboration in the $\psi^{\prime}\pi$ mode in the $B$ decays, which favors the spin-parity
$1^+$ over other hypotheses~\cite{zc4475confirmed}. The radial excited state $[cu][\bar{c}\bar{d}]$ with $3^5D_1$ and $J^P=1^+$ has a mass of
$4497\pm7$ MeV, which is consistent with the energies of the two states. Therefore, the main component of two states can be described as the tetraquark state $[cu][\bar{c}\bar{d}]$ with $3^5D_1$ and $J^P=1^+$ in the CFTM.

In addition, the hidden charmonium pentaquark states $P_c^+(4380)$ and $P_c^+(4450)$ were also investegated in the CFTM~\cite{pc4380&4450}. The
main component of the state $P^+_c(4380)$ can be described as a compact pentaquark state $uudc\bar{c}$ with the pentagonal color structure and $J^P=\frac{3}{2}^-$ in the CFTM. However, the state $P^+_c(4450)$ cannot be accommodated in the CFTM because of the opposite parity with the
experimental result of the state $P^+_c(4380)$.

From the above numerical analysis, it can be found that the most of the hidden charmed states can be matched with the tetraquark states $[qc][\bar{c}\bar{q}]$ in the CFTM. The stability of the tetraquark states can be identified by the binding energies correspond to the meson-meson thresholds $T_{M_1M_2}=M_1(c\bar{q})+M_2(\bar{c}q)$ and $T_{M'_1M'_2}=M_1(c\bar{c})+M_2(q\bar{q})$. The binding energies are defined as $E_b=E_4-T_{M_1M_2}$ and $E_b'=E_4 -T_{M'_1M'_2}$ similar to the research on the stability of tetraquark states ~\cite{threshold}. The states $[cq][\bar{c}\bar{q}]$ with $E_b<0$ and $E_b'<0$ are bound states and cannot decay into two corresponding color singlet mesons under the strong interaction. The other states $[cq][\bar{c}\bar{q}]$ are unstable and can decay into two corresponding color singlet mesons through the rupture and rearrangement of the color flux tubes. It can be found from Table IV that only the states $[cu][\bar{c}\bar{d}]$ with $1^3S_1$ and $1^5S_2$ lie below the thresholds of $D^*\bar{D}$ or $D\bar{D}^*$ and $D^*\bar{D}^*$ by $11$ MeV and $3$ MeV, respectively. So these two states cannot decay into $D^*\bar{D}$ or $D\bar{D}^*$ and $D^*\bar{D}^*$ through strong interactions in the CFTM, respectively. The energies $E_4$ of all of these states are much higher than the corresponding threshold $T_{M'_1M'_2}$ because the large binding energies of the light mesons $\pi$ and $\rho$, which originates from the stronger interactions between two light quarks $q$ and $\bar{q}$. In one word, the tetraquark states $[cq][\bar{c}\bar{q}]$ are impossible to form stable states so that they finally decay into two mesons $c\bar{c}$ and $q\bar{q}$, which is in agreement with the conclusions in many researches on tetraquark states~\cite{threshold,x3872slamanca}. On the contrary, the states $[cc][\bar{q}\bar{q}]$ are easier to form stable tetraquark states beacuse they can only decay into two $c\bar{q}$ mesons in the quark model~\cite{threshold, stable}. The recent discovery of the doubly charmed baryon $\Xi_{cc}$ by the LHCb Collaboration has now provided the crucial experimental input to search for the tetraquark state $[QQ][\bar{q}\bar{q}]$~\cite{xicc}.

The diquark size $\langle\mathbf{r}^2\rangle^{\frac{1}{2}}$, the antidiquark size $\langle\mathbf{R}^2\rangle^{\frac{1}{2}}$ and the distance between the two clusters $\langle\mathbf{X}^2\rangle^{\frac{1}{2}}$ of the charged tetraquark states $[cu][\bar{c}\bar{d}]$ as an example are also calculated and listed in Table V. The diquark and antidiquark are found to share the same size, which is mainly determined by the total spin $S$ but does not change greatly with the total spin $S$, especially for higher orbital excited states. The diquark and antidiquark are therefore rather rigid against the rotation. The distance $\langle\mathbf{X}^2\rangle^{\frac{1}{2}}$ changes remarkably with the relative orbital excitation $L$ between the two clusters and however is irrelevant to the total spin $S$. In this way, one can figure out the picture that the diquark and the antidiquark look like very compact objects well separated one from each other, which was called as dumbbell configuration in the work on the properties of diquonia~\cite{dumbbell}. The higher the orbital angular momentum $L$, the more prolate the shape of the excited states. The three-dimension spatial configuration is determined by the dynamics of the model, especially the multibody confinement potential and kinetic energy. The color flux tubes reduce the distance between any two connected quarks to as short a distance as possible to minimize the confinement potential energy, while the kinetic motion expands the distance between any two quarks to as long a distance as possible to minimize the kinetic energy. Therefore, they compete with each other to eventually achieve an optimum spatial structure: three dimensional compact structure. LQCD study on the tetraquark states demonstrated that the twisted tetraquark configuration or the tetrahedral structure seems to be rather stable against the
transition into the two mesons~\cite{lqcd2}.
\begin{table}
\caption{The rms $\langle\mathbf{r}^2\rangle^{\frac{1}{2}}$, $\langle\mathbf{R}^2\rangle^{\frac{1}{2}}$ and
$\langle\mathbf{X}^2\rangle^{\frac{1}{2}}$ of charged tetraquark states $[cu][\bar{c}\bar{d}]$ with $S$ and $L$, unit in fm.}
\begin{tabular}{ccccccccccccccccccccc}

\hline
         ~~~~$SL$~~~~                      & ~~$00$~~ & ~~$01$~~ & ~~$02$~~ & ~~$03$~~ & ~~$10$~~ & ~~$11$~~ & ~~$12$~~ & ~~$13$~~  \\
$\langle\mathbf{r}^2\rangle^{\frac{1}{2}}$ &   0.85   &   0.94   &   0.95   &   0.96   &   0.90   &   0.96   &   0.98   &    0.99   \\
$\langle\mathbf{R}^2\rangle^{\frac{1}{2}}$ &   0.85   &   0.94   &   0.95   &   0.96   &   0.90   &   0.96   &   0.98   &    0.99   \\
$\langle\mathbf{X}^2\rangle^{\frac{1}{2}}$ &   0.42   &   0.85   &   1.09   &   1.27   &   0.48   &   0.85   &   1.10   &    1.30   \\
             $SL$                          & ~~$20$~~ & ~~$21$~~ & ~~$22$~~ & ~~$23$~~ \\
$\langle\mathbf{r}^2\rangle^{\frac{1}{2}}$ &   0.98   &   1.00   &   1.01   &   1.02   \\
$\langle\mathbf{R}^2\rangle^{\frac{1}{2}}$ &   0.98   &   1.00   &   1.01   &   1.02   \\
$\langle\mathbf{X}^2\rangle^{\frac{1}{2}}$ &   0.57   &   0.92   &   1.12   &   1.30   \\

\hline
\end{tabular}
\end{table}

The multibody confinement potential based on the color flux-tube picture is the dynamical mechanism of the formation of the compact tetraquark states, which is a collective degree of freedom and binds all particles to establish a compact multiquark state. The higher $L$ the excited state, the more confinement potential is stored in the color flux tube connecting the diquark and antidiquark because the confinement is proportional to the distance between the two clusters $\langle\mathbf{X}^2\rangle^{\frac{1}{2}}$. In the case of well-defined tetraquark states, it can be found in Table IV that the experimental decay width of the charged states $Z_c^+$ is really proportional to $L$, which is supported by the theoretical investigations on the decay width of the mesons and tetraquark states in the string model~\cite{decaywidth}. This phenomenon does not seem obvious for the $XY$ states, which originate from that the main component of a few or some of $XY$ states may be not diquark-antiquark states but meson-meson molecule states, charmonium or hybrids, et al.

The assignment of the diquark-antidiquark component of the hidden charmed states in the CFTM is completed just according to the proximity to the experimental masses. As a matter of fact, a tetraquark system should be the mixture of the diquark-antidiquark and meson-meson configurations,
which represents an interesting phenomenon of the flip-flop, namely a recombination of the color flux-tube configuration so as to minimize the total confinement potential in accordance with the change of the quark location. The flip-flop is important for the properties of tetraquark states especially for the decay process into two mesons. The hidden charmed states should eventually decay into several color singlet mesons due to their high energy. In the course of the decay, the three-dimension spatial structure must collapse because of the breakdown of the color flux tubes, and then the decay products form by means of the recombination of color flux tubes. The decay widths of the hidden charmed states are determined by the transition probability of the breakdown and recombination of color flux tubes, which is worthy of further research to inspect strictly the main component of the hidden charmed states in future work. In addition, the flip-flop leads to infrared screening of the long-range color interactions between two particles in different mesons, and so that the color van der Waals force between two mesons disappear~\cite{lqcd2}.

\section{summary}

We systematically investigate the hidden charmed states observed in experiments within the framework of the CFTM involving a multibody confinement potential instead of a two-body one  proportional to the color charge in the traditional quark models. Our model investigations demonstrate that the most of the states can be universally identified as compact tetraquark states just taking the proximity to the experimental masses into account. The stringent check of the assignment of the main component of the hidden charmed states is indispensable by systematically investigating on the decay properties of the states.

These discoveries of multi-quark hadrons, at least the charged states $Z_c$, have revealed new aspects of hadron physics, especially for the complicated non-perturbative behavior of QCD. The multibody color flux-tube in the multi-quark states employs a collective degree of freedom whose dynamics play an important role in the formation and decay of those compact states.

In the present calculation, only diquark-antidiquark configuration are considered. Di-meson structure is also possible and should be taken into account by introducing the flip-flop confinement potential. Furthermore, for the most states, the quark-antiquark configuration can not be ruled out, the mixing of the quark-antiquark with the tetraquark states will move the physical states up or down. Therefore, the more complete calculation includes all the effects are expected to give a more reliable description of these hadron states.

\acknowledgments
{This research is partly supported by the National Science Foundation of China under Contracts Nos. 11775118, 11675080, 11535005 and Fundamental Research Funds for the Central Universities under Contracts No. SWU118111.}

\end{document}